\documentclass[prb,twocolumn,amsmath,amssymb]{revtex4}

\usepackage[dvips]{graphicx}
\usepackage{bm}
\usepackage{times}

\begin{document}

\title{
Quantum Interference Controlled Molecular Electronics
}

\author{San-Huang Ke,$^{1}$ Weitao Yang,$^{1}$ and Harold U. Baranger$^{2}$}

\affiliation{
     $^{\rm 1}$Department of Chemistry, Duke University, Durham, NC 27708-0354 \\
     $^{\rm 2}$Department of Physics, Duke University, Durham, NC 27708-0305
}

\date{August 29, 2008; published as Nano Lett.\ 8, 3257-3261 (2008)}

\begin{abstract}
Quantum interference in coherent transport through single molecular
rings may provide a mechanism to control current in molecular
electronics. We investigate its applicability by using a single-particle
Green function method combined with {\it ab initio} electronic structure
calculations. We find that the quantum interference effect (QIE) depends
strongly on the interaction between molecular $\pi$ states and contact
$\sigma$ states. It is masked by $\sigma$ tunneling in small molecular
rings with Au leads, such as benzene, due to strong $\pi$-$\sigma$
hybridization, while it is preserved in large rings, such as
[18]annulene, which then could be used to realize QIE transistors. 
\end{abstract}

\maketitle

The use of single molecules as functional devices is the ultimate end of the ongoing trend toward miniaturization of electronic circuits \cite{NitzanRatner03review,Tao06review,Koentopp08083203}. Despite significant progress made in the last decade, several issues still challenge the realization of molecular electronics, notably the sensitivity and control of molecule-lead contacts \cite{Venkataraman06458,Basch051668,Ke05074704}. Recently Cardamone et al.\ \cite{Cardamone062422,Stafford07424014} proposed a novel mechanism to control electron transport through single molecular rings: the current is determined by the degree of destructive or constructive quantum interference between the two paths around a symmetric molecule. Such interference can be controlled by a third terminal providing elastic scattering or dephasing. Their use of a quantum interference effect (QIE) in the completely coherent quantum regime builds on previous work in semiconductor nanostructures \cite{Sols89350,Goodnick031264896} and molecular nanostructures \cite{Sautet88511,Baer024200,Yaliraki02153,Hettler03076805,Stadler03138,Stadler05S155}. Since the QIE stems essentially from the symmetry of a molecular device, it should not be affected significantly by the structure of the molecule-lead contact. We present the first study of this novel mechanism in realistic \emph{ab initio} calculations. 

QIE controlled molecular electronics was proposed \cite{Cardamone062422,Stafford07424014} using a model calculation for a benzene ring in which only the $\pi$ molecular states were considered. The connection between the leads and the molecular states was treated phenomenologically, and the possibility of induced structural relaxation was ignored. As these simplifications could have a big effect on the conductance, it is critical to investigate the validity of the QIE mechanism in realistic systems and to evaluate the factors affecting it.

In this paper, we investigate the QIE mechanism by performing quantum transport calculations for two molecular rings, one small -- benzene -- and one large -- [18]annulene. We adopt a standard single-particle Green function method \cite{Haug96,Ke04085410} combined with {\it ab initio} electronic structure calculations in which all the above-mentioned factors are fully taken into account. Two different electrodes (leads) are studied: gold, which has $\pi$-$\sigma$ molecule-lead coupling, and a metallic (5,5) carbon nanotube (CNT), which has strong $\pi$-$\pi$ coupling.

Our calculations show that the survival of the QIE mechanism depends
strongly on the interaction between the molecular $\pi$ states and the
contact $\sigma$ states. It is masked by $\sigma$ tunneling in the small
benzene/Au system because of the strong $\pi$-$\sigma$ hybridization,
while it is preserved in the large [18]annulene/Au system, which then
could be used to realize QIE transistors. With the CNT leads, we find
that strong $\pi$-$\pi$ molecule-lead coupling can modify significantly
the QIE but does not destroy it. 

Our method for calculating electron transport through molecular junctions has been described in detail previously \cite{Ke04085410}; it combines a Landauer approach to transport with {\it ab initio} electronic structure methods \cite{DiVentra00979,Taylor01245407,Damle01201403,Xue02151,Brandbyge02165401, Louis03155321,Rocha06085414,Arnold07174101}. 
In particular, we use here density functional theory (DFT) \cite{Parr89} in the local density approximation (LDA), hybrid DFT in the version of B3LYP \cite{Becke935648,Lee785}, and Hartree-Fock (HF) theory. Several energy functionals are used to make sure that the conclusions are generally valid, because to date no functional is completely accurate for transport calculations. Usually DFT-LDA and HF are the two extremes, the former overestimates and the latter underestimates the conductance \cite{Ke07201102}. The wave functions are expanded using a Gaussian 6-311G** basis set for the C, H, and S atoms, and a CRENBS basis set for Au atoms \cite{NWChem}.

Practically, the lead-molecule-lead system is divided into three parts:
left lead, right lead, and device region. The latter contains the
molecule plus parts of the leads to accommodate the molecule-lead
interaction (see structures below). The self-consistent DFT or HF
Hamiltonian of the device region plus the self-energies of the two
semi-infinite leads, ${\bf \Sigma}_{L,R}(E)$, are used to construct a
single-particle Green function, ${\bf G}_D(E)$, from which the
transmission coefficient as a function of energy is calculated:
$T(E)={\rm Tr}[{\bf \Gamma}_L{\bf G}_D{\bf \Gamma}_R{\bf G}_D^{\dag}]$
where ${\bf \Gamma}_{L,R}(E) = i[{\bf \Sigma}_{L,R}(E) - {\bf
\Sigma}^{\dag}_{L,R}(E)]$ is the coupling to the left or right lead. The
conductance, $G$, then follows from a Landauer-type relation. 

For the benzene ring, we consider only gold leads. The self-energy due
to the Au within the leads is treated in the wide
band limit (WBL) approximation: ${\bf \Gamma}(E) = -\gamma {\bf I}$ with
$\gamma$ = 3.0 eV \cite{wbl-note}. But note that the key molecule-lead
coupling is included explicitly in the Hamiltonian because 9
Au atoms are included in the central device region. Since the density of
states of gold is quite flat around the Fermi energy, the use of the WBL
for Au atoms is reasonable. For the large
[18]annulene ring, we consider gold leads as for benzene as well as
(5,5) metallic carbon nanotube leads for which the self-energy is
obtained by {\it ab initio} calculation of atomic leads. For all systems
the atomic structure of the junction, including the molecule-lead
separation, is optimized \cite{DZP-optimization} by minimizing the atomic forces on atoms to be smaller than 0.02 eV/{\AA}.

\begin{figure}
(a)\includegraphics[angle= 90, height=3.5cm]{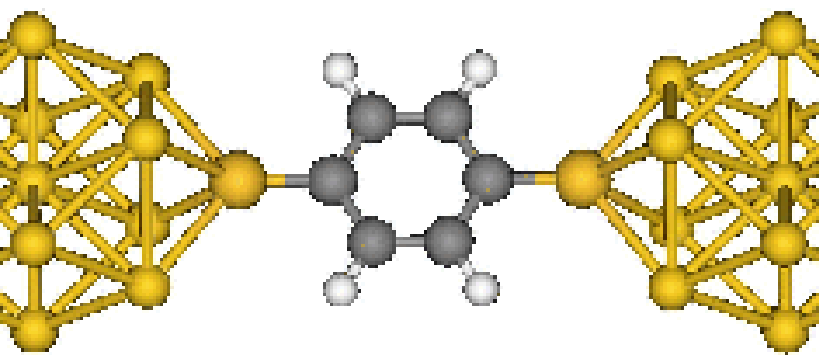}
(b)\includegraphics[angle=   0,height=3.5cm]{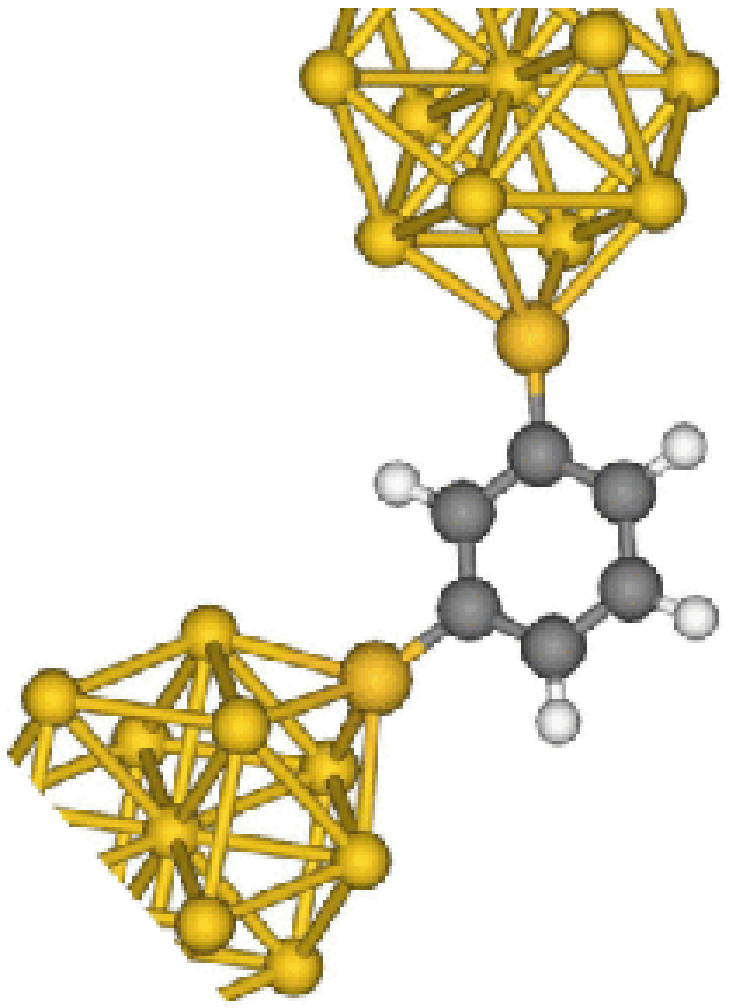}
(c)\includegraphics[angle=   0,height=3.5cm]{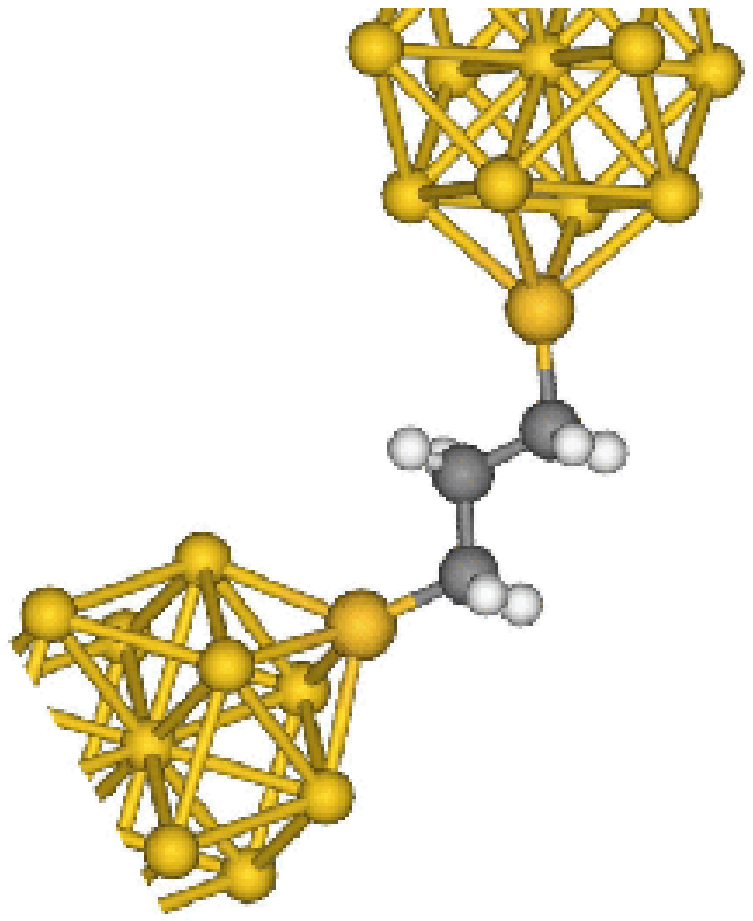}
(d)\includegraphics[angle=-90,width=7.5cm]{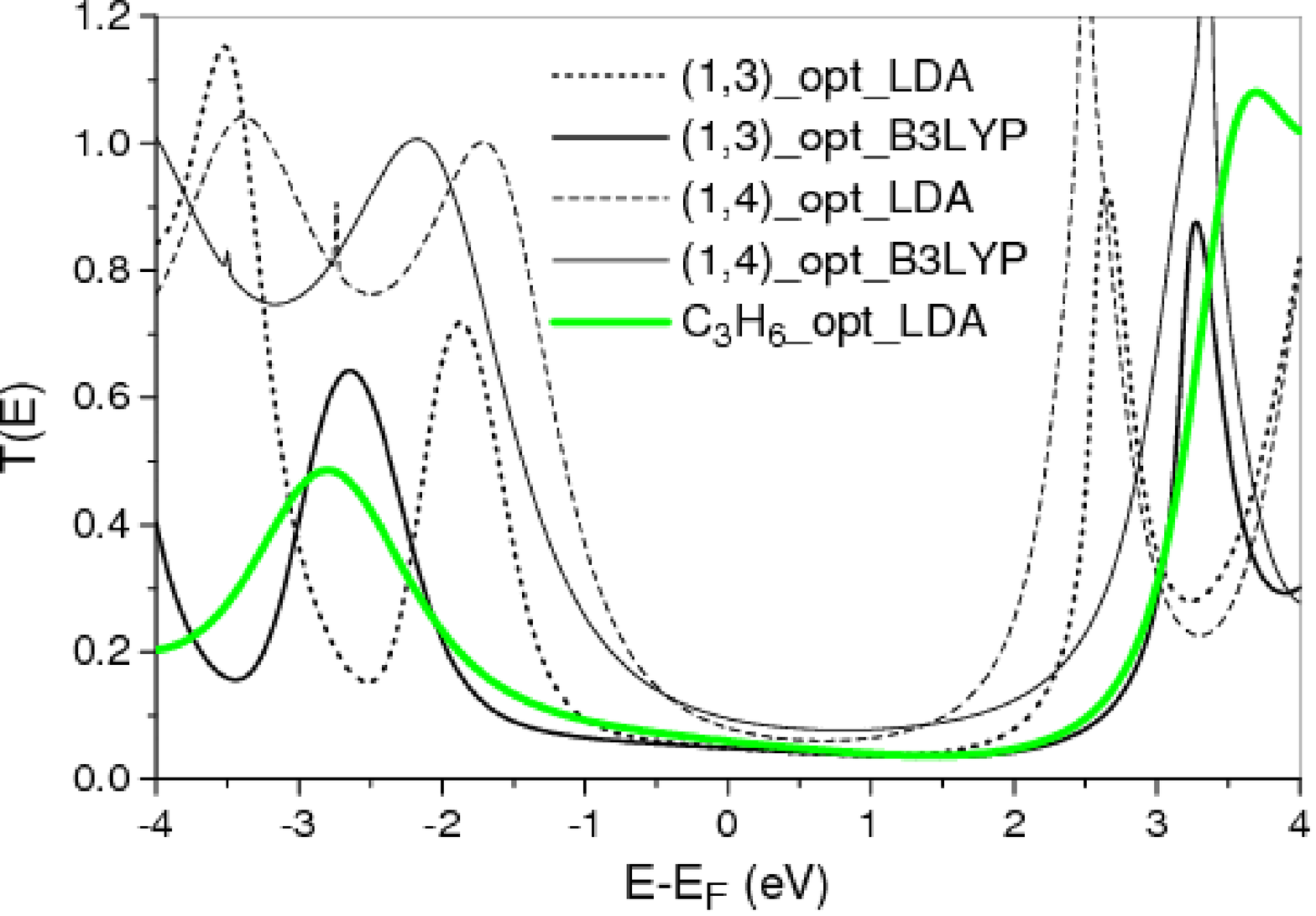}
(e)\includegraphics[angle=-90,width=7.5cm]{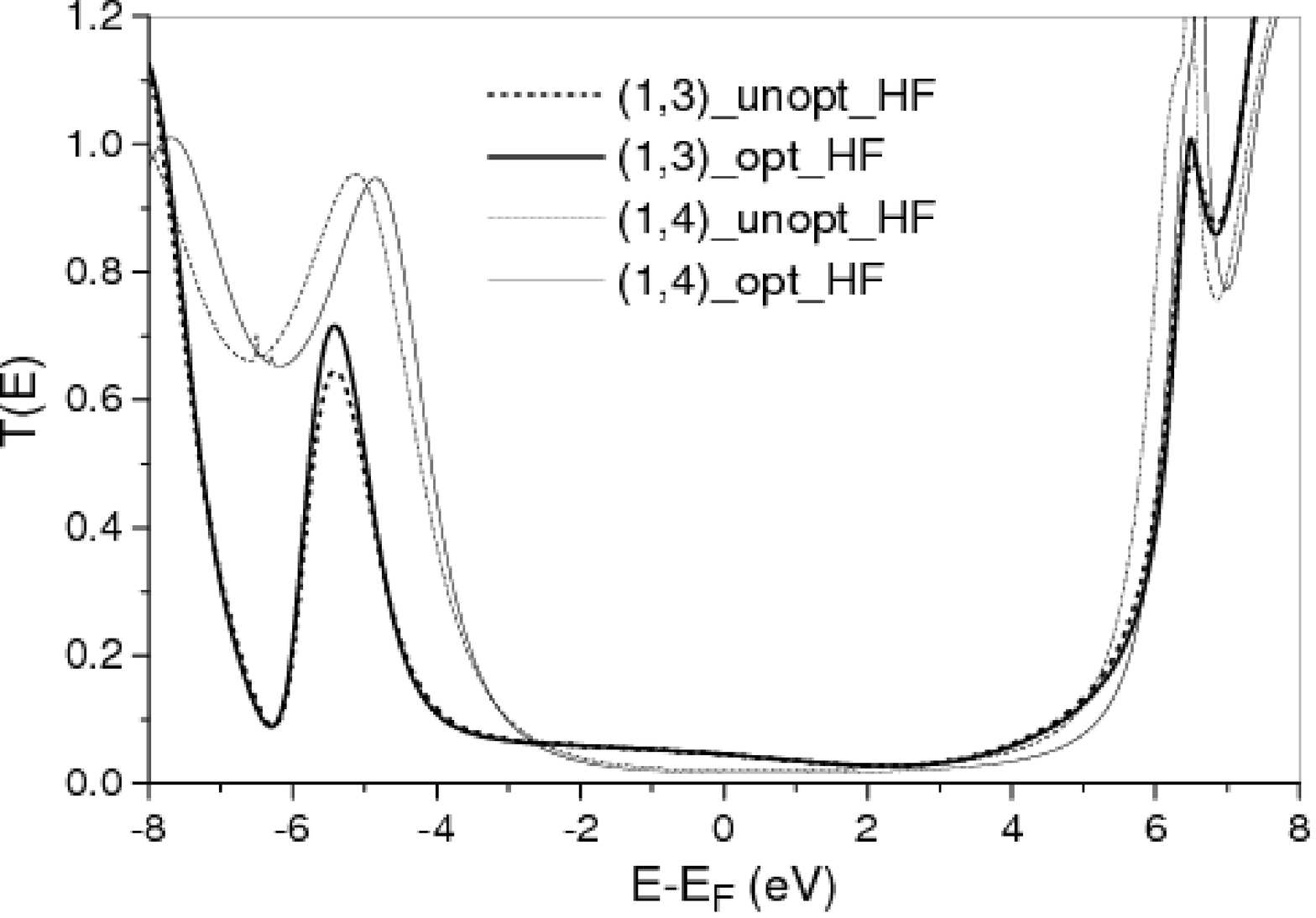}
\caption{(color online) Optimized structures of the Au-S-benzene-S-Au
system (device region only) with (a) the (1,4) constructive and (b) the
(1,3) destructive interference configurations. (c) The alkane
configuration used for comparison. (d),(e) Transmission functions
obtained with different energy functionals. For the HF calculation we
consider cases both with and without the lead-induced molecular
relaxation, as indicated. Note the absence of a clear difference between
the (1,3) and (1,4) cases. 
}
\label{fig_str_bdt}
\end{figure}


Let us start with the small system -- the benzene ring in Fig.~1(a). All of the Au atoms included in the device region are shown; the WBL self-energy is applied to these atoms. The S atom is situated on a hollow site. Two structural configurations are considered: (1,4) and (1,3) in Figs.\ 1(a) and 1(b), respectively. Configuration (1,4) has constructive interference in the all-$\pi$ model (phase difference between the two paths around the ring is 0), while (1,3) has destructive interference (phase difference is $\pi$) \cite{Cardamone062422,Stafford07424014}. Still within the all-$\pi$ model, the non-equilibrium many-body physics was subsequently studied in detail  for these two configurations \cite{Begemann08201406}.

The transmission $T(E)$ found using several functionals is shown in
Figs.\ 1(d) and (e). First, note that the transmission gap depends on
the functional as expected: it increases in the order LDA, B3LYP, and
HF. Also, the LDA equilibrium conductance for the (1,4) configuration is
consistent with previous calculations
\cite{Xue014292,Xue03115407,Ke05114701}, about $0.1 G_0$. 
The key result here is the comparison between the (1,4) and the (1,3)
configurations. The all-$\pi$ model predicts that the (1,3)
configuration has a transmission node at the Fermi energy due to total
destructive interference \cite{Cardamone062422,Stafford07424014}. However, in our more
realistic calculation, this transmission node does not occur. For LDA
and B3LYP $G_{(1,3)}$ is smaller than $G_{(1,4)}$ by a factor of about
2; for HF the order is even reversed, $G_{(1,3)} > G_{(1,4)}$. 

Several factors may destroy the perfect destructive interference 
present in the model of Refs.\ \onlinecite{Cardamone062422} and \onlinecite{Stafford07424014}:
(i) lead-induced structural relaxation which breaks the symmetry, 
(ii) the influence of the $\sigma$ states, and 
(iii) beyond nearest-neighbor interactions which cause other paths. 
Let us first look at the effect of structural relaxation using the HF
calculation: results for relaxed and unrelaxed structures are compared
in Fig.\ 1(e). The effect on $T(E)$ is clearly minor. Because of the
symmetry of the additional paths, the effect of (iii) should also be
small. Thus, we conclude that factors (i) and (iii) are not important. 

\begin{figure}[tb]
(a)\includegraphics[angle= 0,height=3.0cm]{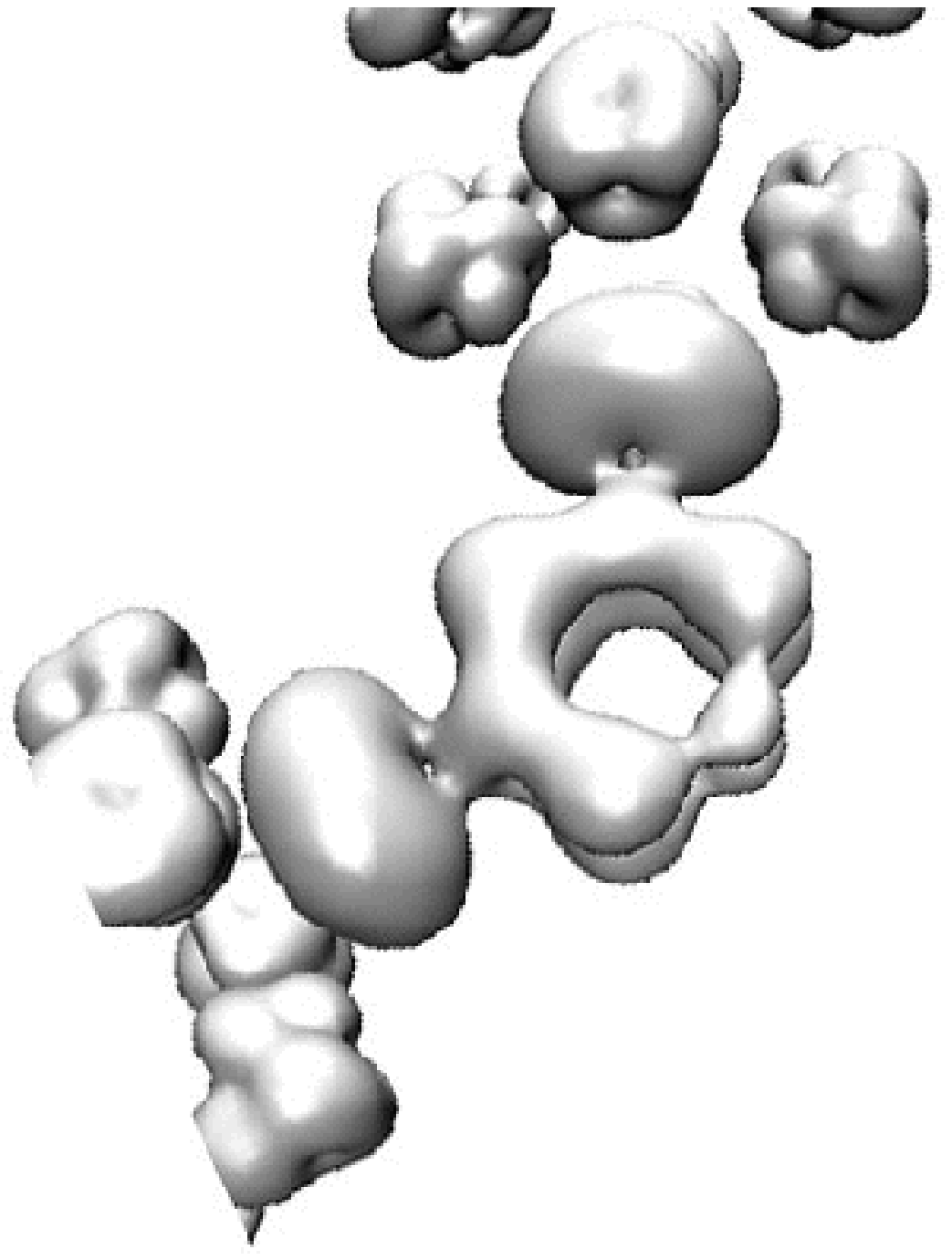}
(b)\includegraphics[angle= 0,height=3.0cm]{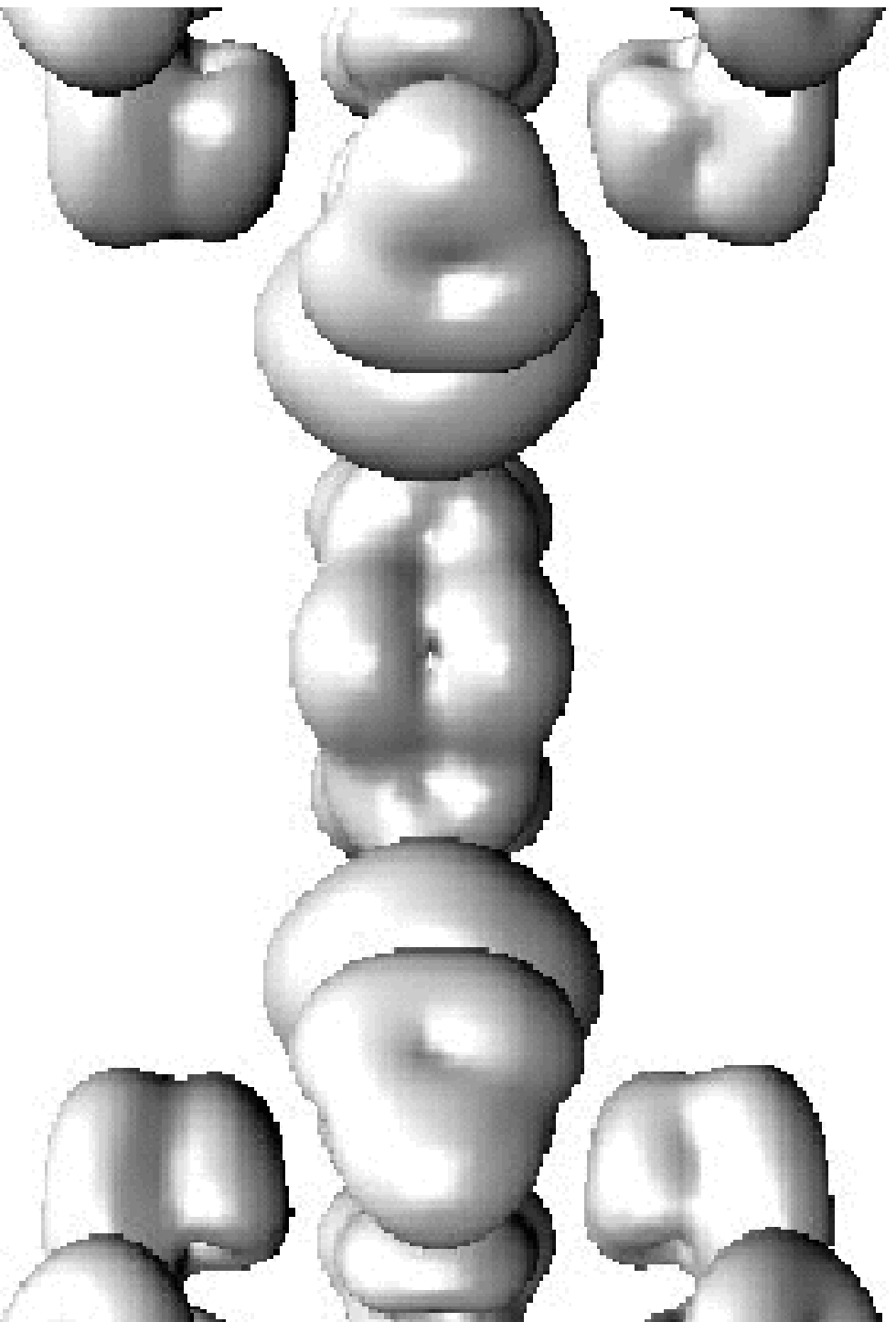}
(c)\includegraphics[angle= 0,height=3.0cm]{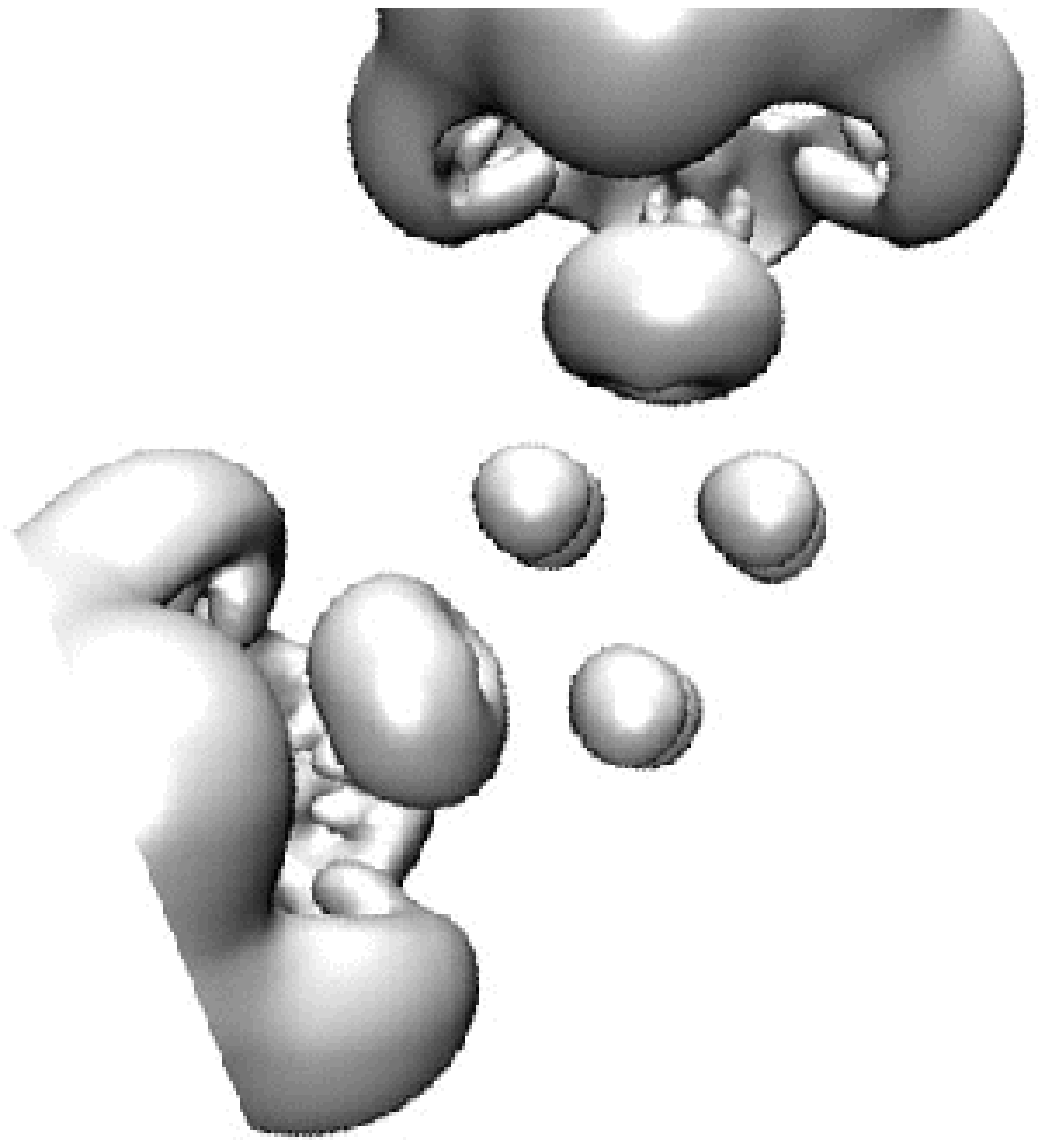}
\caption{
Local density of states (LDOS) for the (1,3) benzene/Au system using
LDA. (a) Top view and (b) side view for the energy window [-2.5, -1.0]
eV (HOMO resonance) in Fig.\ 1 (d). (c) Top view for the energy window
[-0.2, 0.2] eV. The influence of the $\sigma$ bonds is clear in (a) and
(b).  
}
\label{fig_LDOS_BDT}
\end{figure}

\begin{figure}[tb]
(a)\includegraphics[angle=  0,height=2.5cm]{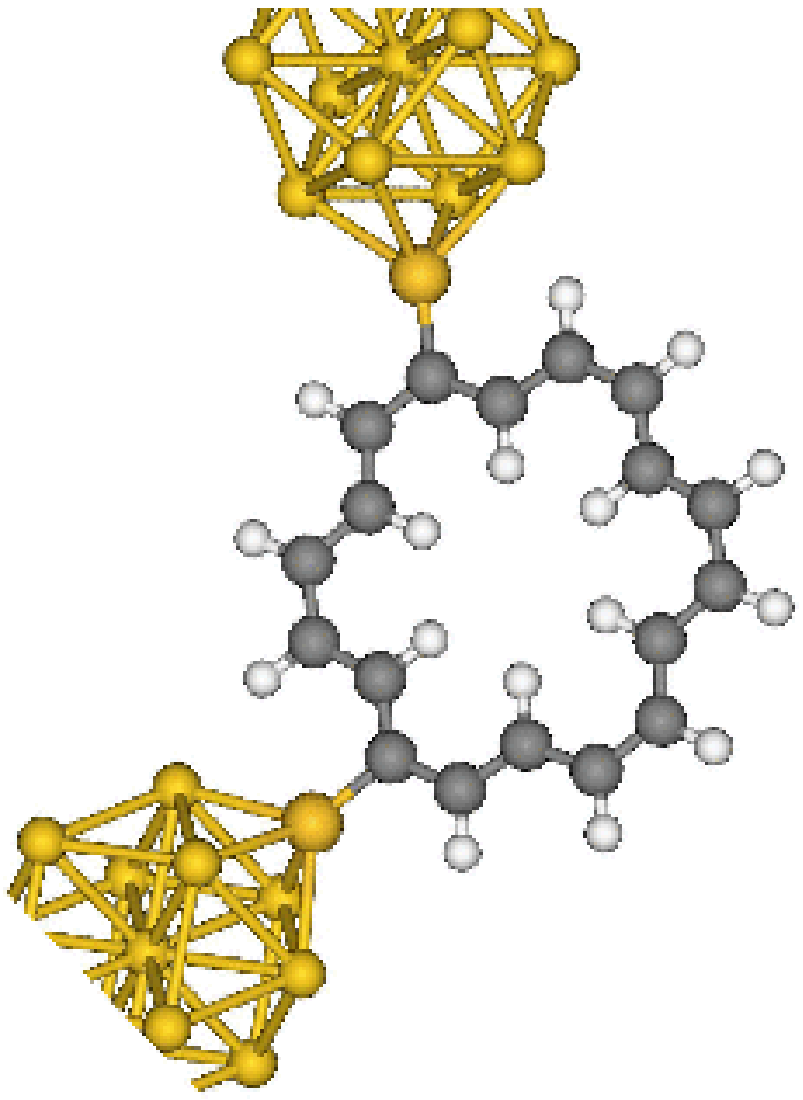}
(b)\includegraphics[angle=  0,height=2.5cm]{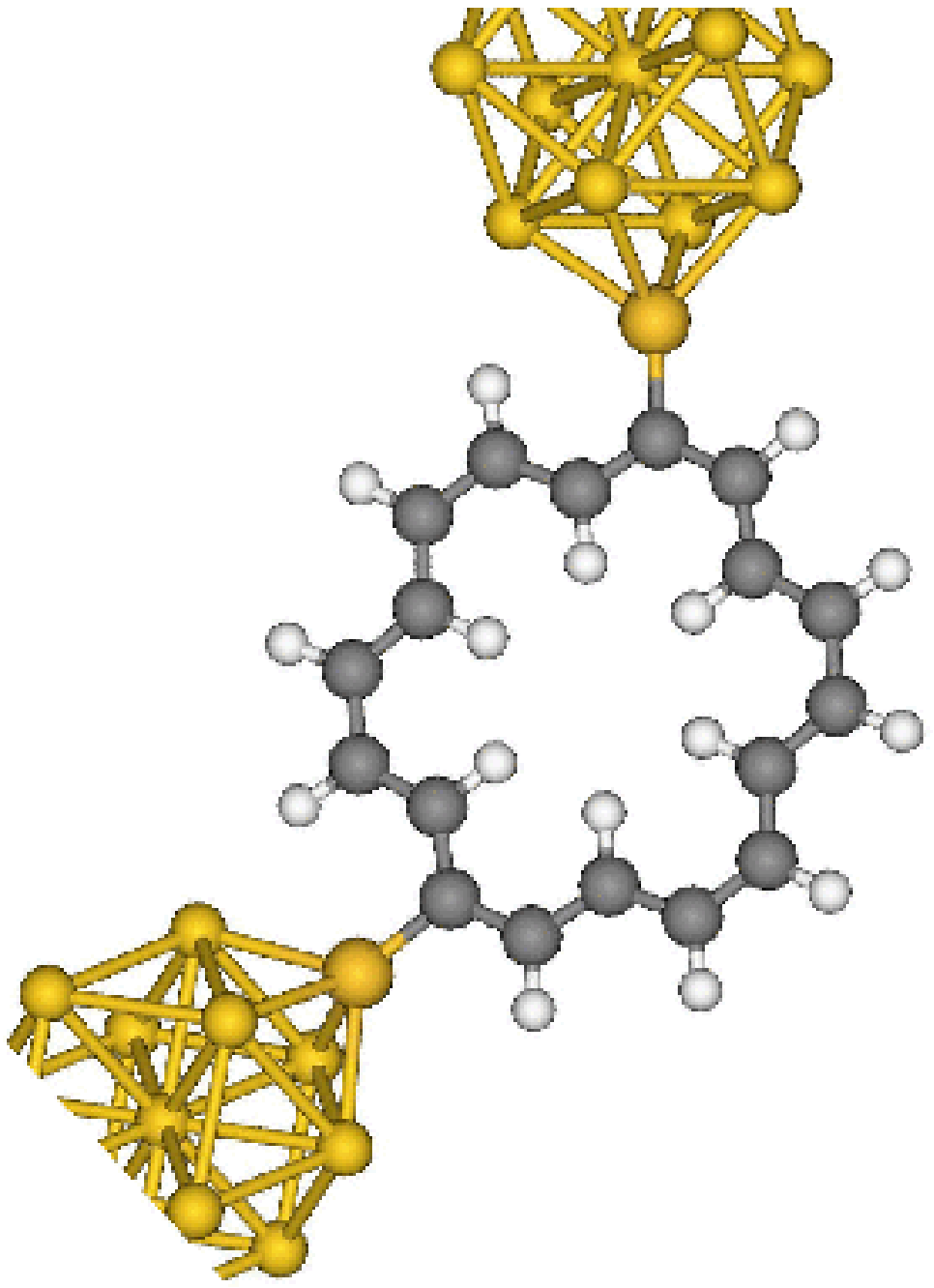}
(c)\includegraphics[angle= 90,height=2.5cm]{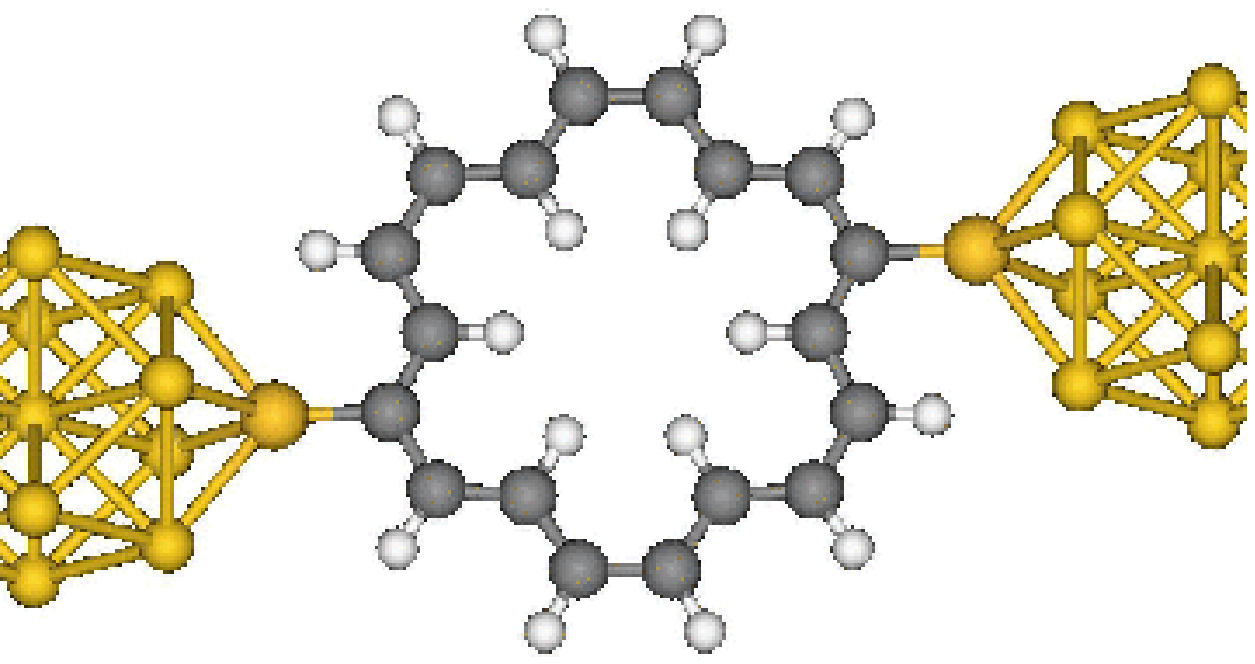}
(d)\includegraphics[angle=  0,height=2.5cm]{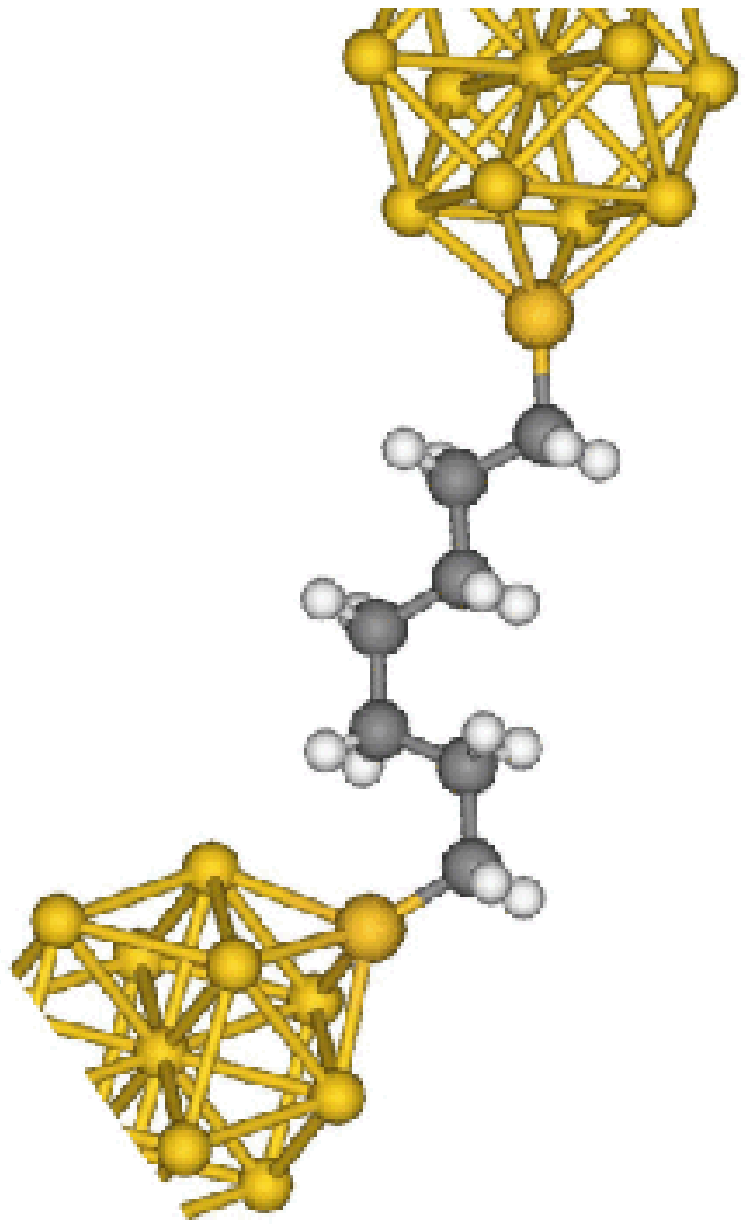} \\
\vspace*{3mm}
(e)\includegraphics[angle=-0,width=7.0cm]{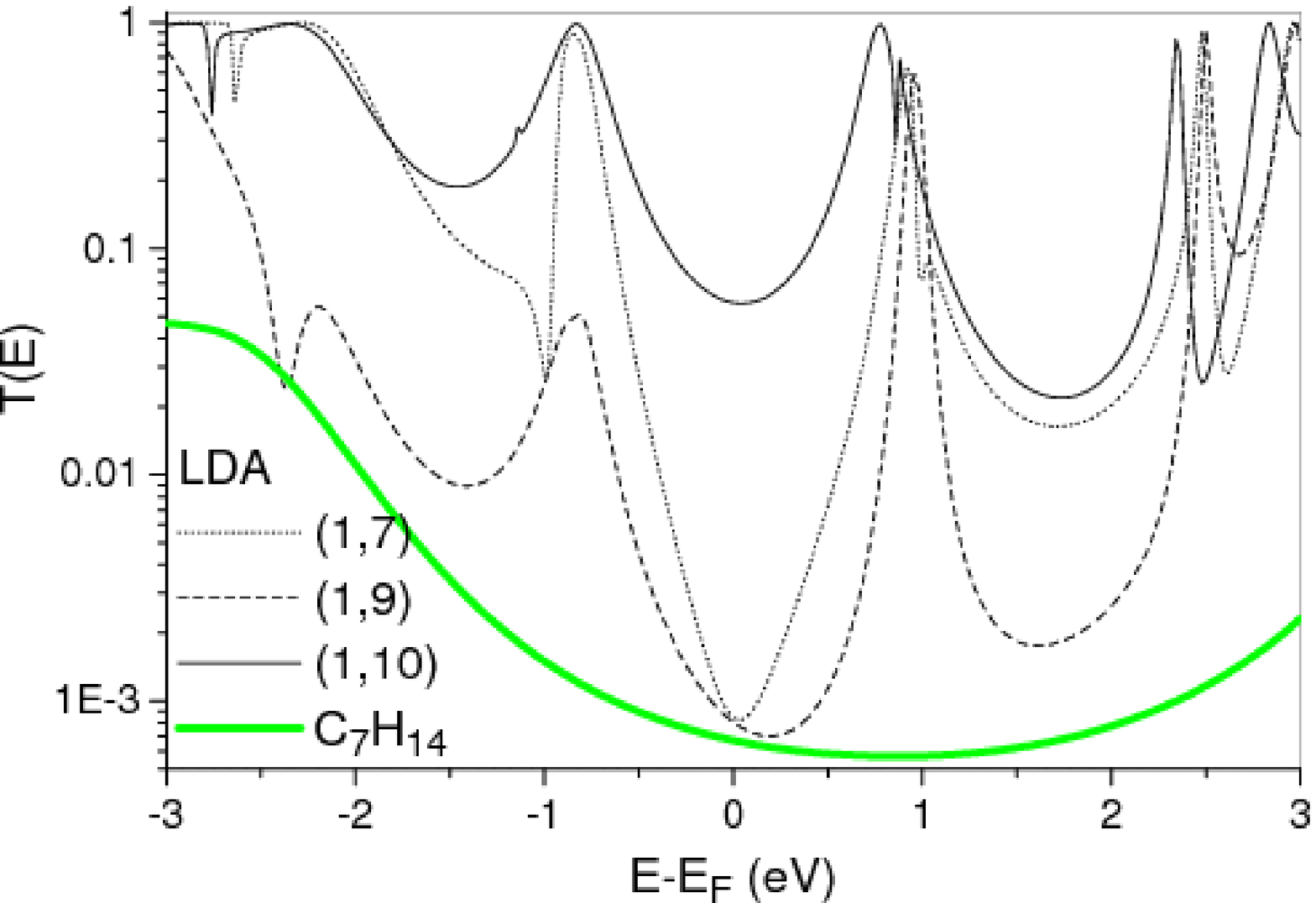}
(f)\includegraphics[angle=-0,width=7.0cm]{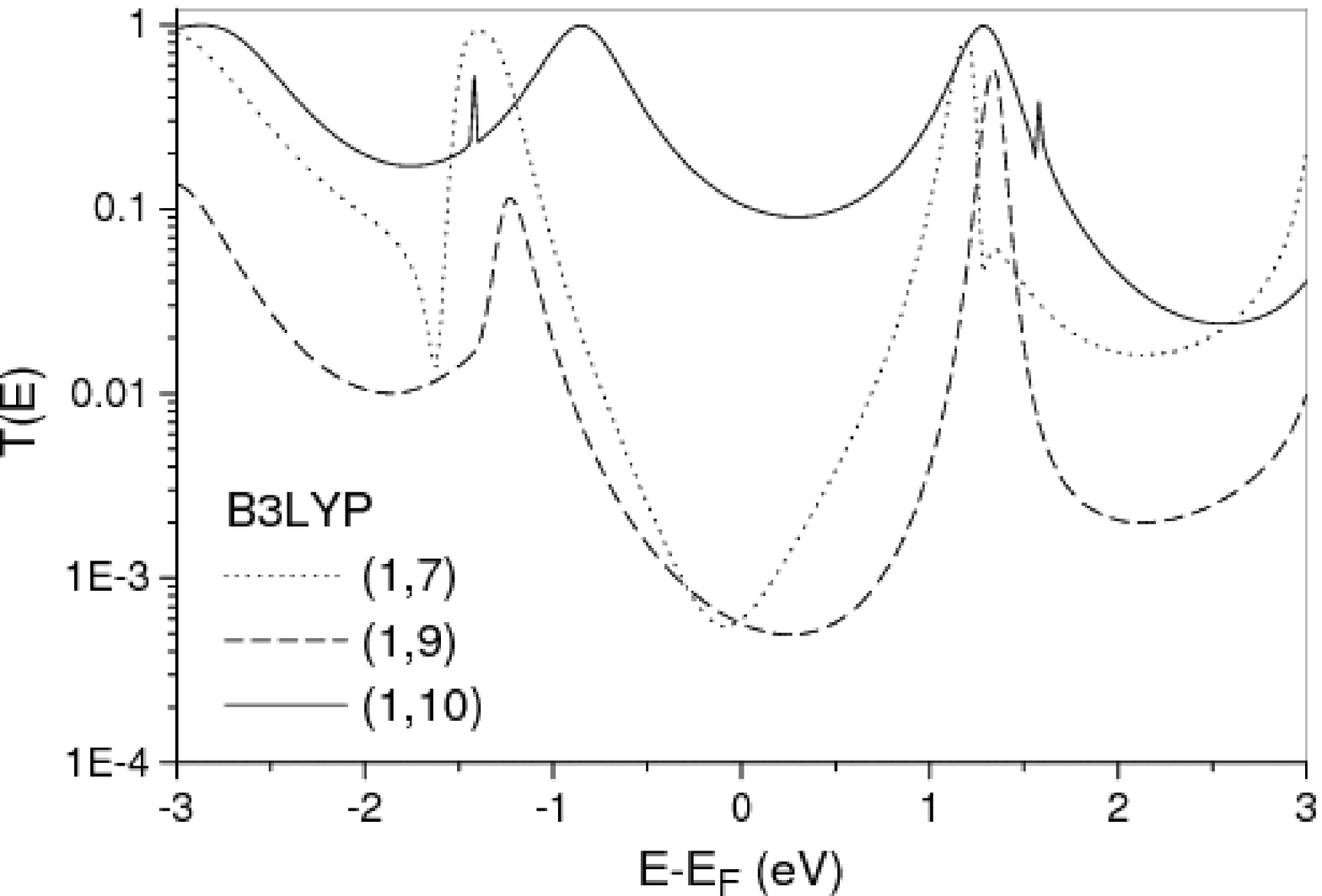}
(g)\includegraphics[angle=-0,width=7.0cm]{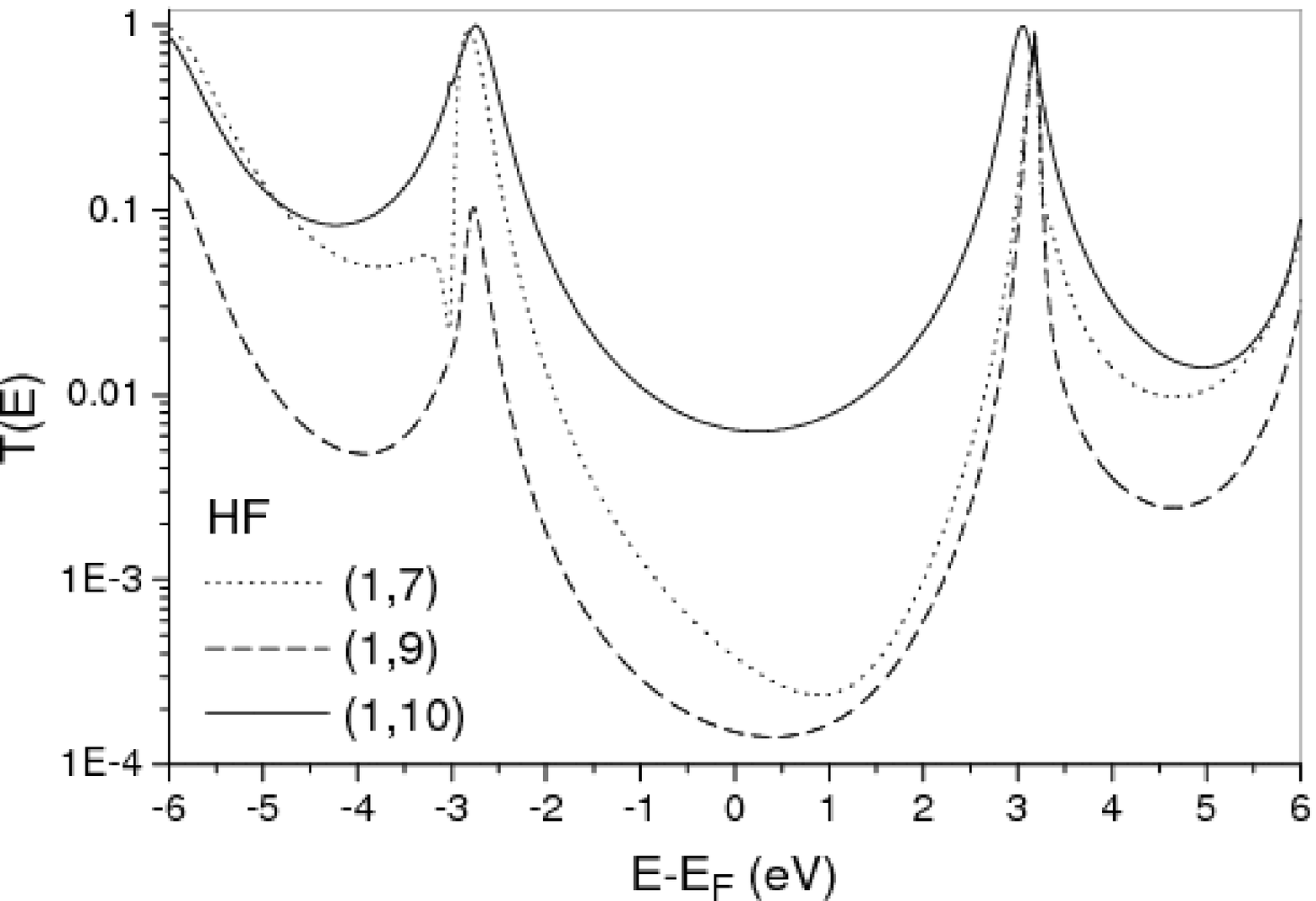}
\caption{
(color online) Optimized structures for Au-S-[18]annulene-S-Au (device
region only) with (a) (1,7), (b) (1,9), and (c) (1,10) lead 
configurations. (d) The alkane structure used for comparison.
Transmission functions are given in (e) LDA, (f) B3LYP, and (g) HF.
Note the clear difference between configurations with destructive and
constructive interference.
}
\label{fig_str_ann18}
\end{figure}

The first hint of the strong effect of $\sigma$ hybridization comes from
the difference between the transport gap in Fig.\ 1 and the HOMO-LUMO
gap of an isolated benzene molecule. In LDA, for instance, the former is
$\sim$4.1 eV while the latter is 5.2 eV. This suggests that the HOMO and
LUMO resonances may not be dominated by benzene $\pi$ states. To check
this, we plot in Fig.\ 2 the LDA local density of states (LDOS) both for the
HOMO resonance ($-2.5$ eV to $-1.0$ eV) and for near the Fermi energy
($-0.2$ eV to $0.2$ eV). Figs.\ 2(a) and (b) show that the HOMO
resonance has large contributions from the contact Au-S states. Even in
the benzene ring the $\sigma$ character is remarkable. Because of this
strong $\pi$-$\sigma$ hybridization, transport near $E_F$ consists
largely of tunneling through the $\sigma$ states rather than through the
$\pi$-ring, as clearly shown by the LDOS around $E_F$ in Fig.\ 2(c). To
further examine this conclusion, we replace the benzene ring with a
$\sigma$-bonded C$_3$H$_6$ molecule [Fig.\ 1(c)]. $T(E)$ in Fig.\ 1(d)
(green line) shows that the conductance is very close to that of the
(1,3) system, suggesting a similar transport mechanism.
This shows directly that the conductance in the $(1,3)$ configuration mainly comes from $\sigma$
tunneling which masks the QIE effect.

Next we turn to the larger [18]annulene ring shown in Fig.\ 3. We
consider two configurations which yield destructive interference in the
$\pi$-only model \cite{Cardamone062422,Stafford07424014}, (1,7) and (1,9), and one
constructive configuration, (1,10). From the $T(E)$ given by 
LDA, B3LYP, and HF
one sees the same trend
in the transport gap and in the equilibrium conductance:
the smallest conductance is given by HF
with the largest gap, which is about one order of magnitude
smaller than the LDA result, in agreement with a previous result for
other long-chain molecules \cite{Ke07201102}.
The key feature here is that the conductance for (1,7) and (1,9) is much
smaller than that for the (1,10), by about two orders of magnitude. \textit{This
indicates that, in contrast to benzene, the ideal destructive
interference is largely preserved.} 

To understand this, we show in Fig.\ 4 the LDA LDOS near both the HOMO resonance (-1.0 eV to -0.2 eV) and the Fermi energy (-0.2 eV to 0.2 eV). One sees that the HOMO resonance is dominated by the [18]annulene $\pi$ states, and transport near the Fermi energy is basically through the $\pi$-ring. Because of the near perfect destructive interference of the $\pi$ transmission \cite{Cardamone062422,Stafford07424014}, the residual conductance should come from $\sigma$-bond tunneling. To check this, we replace the [18]annulene ring with a $\sigma$-bonded C$_7$H$_{14}$ chain [see Fig.\ 3(d)]. The $T(E)$ [green line in Fig.\ 3(e)] yields a conductance very close to that of (1,7), confirming our view.

\begin{figure}[tb]
(a)\includegraphics[angle= 0,height=3.3cm]{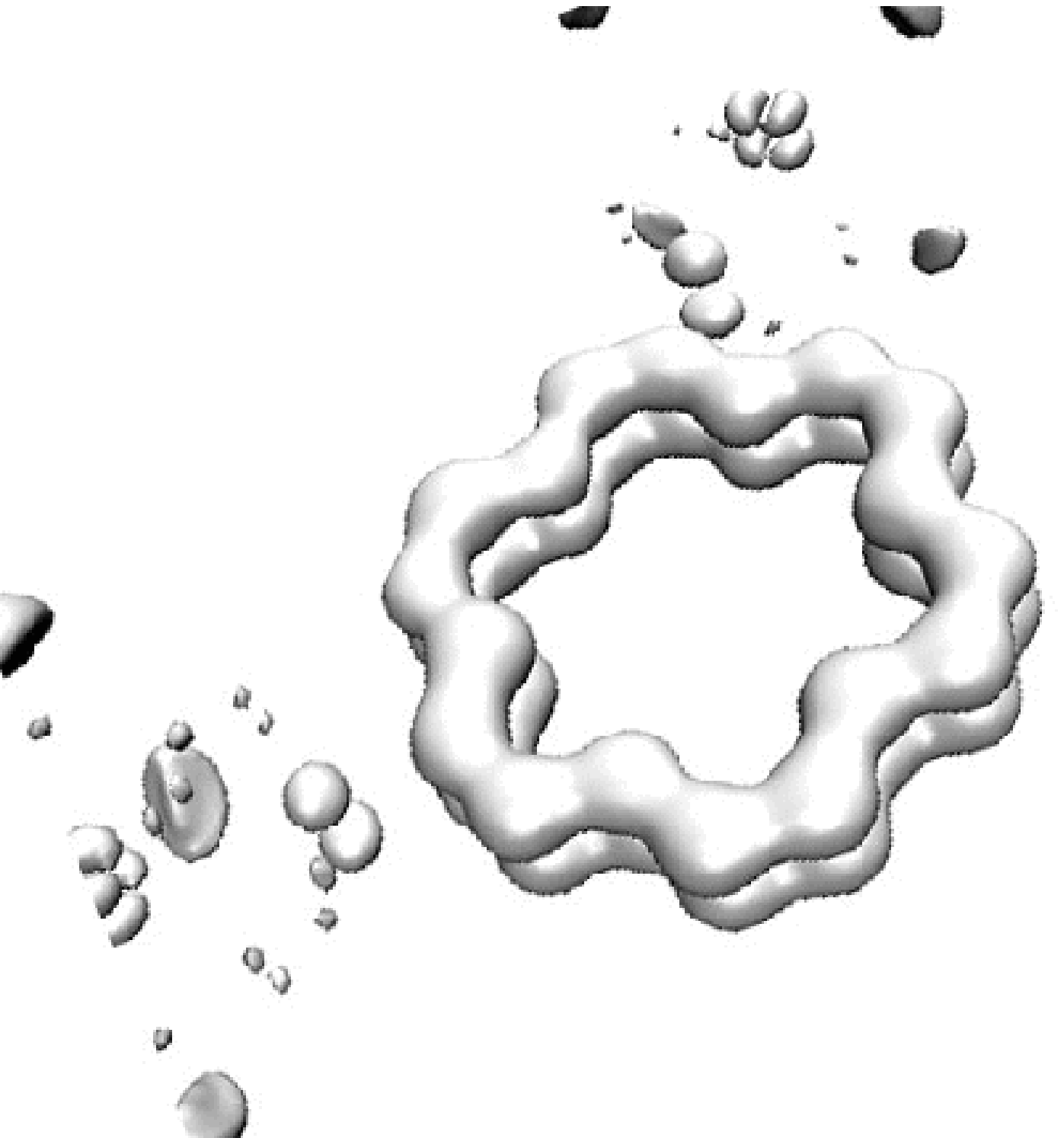}
(b)\includegraphics[angle= 0,height=3.3cm]{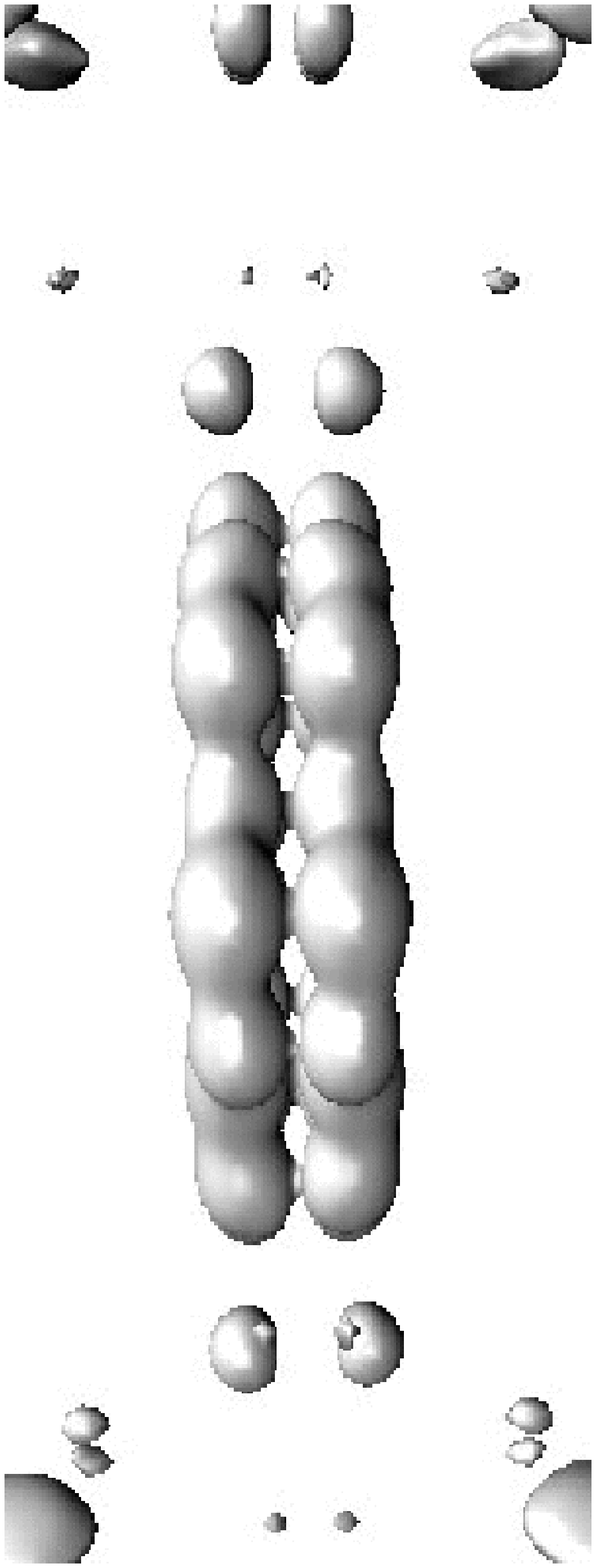}
(c)\includegraphics[angle= 0,height=3.3cm]{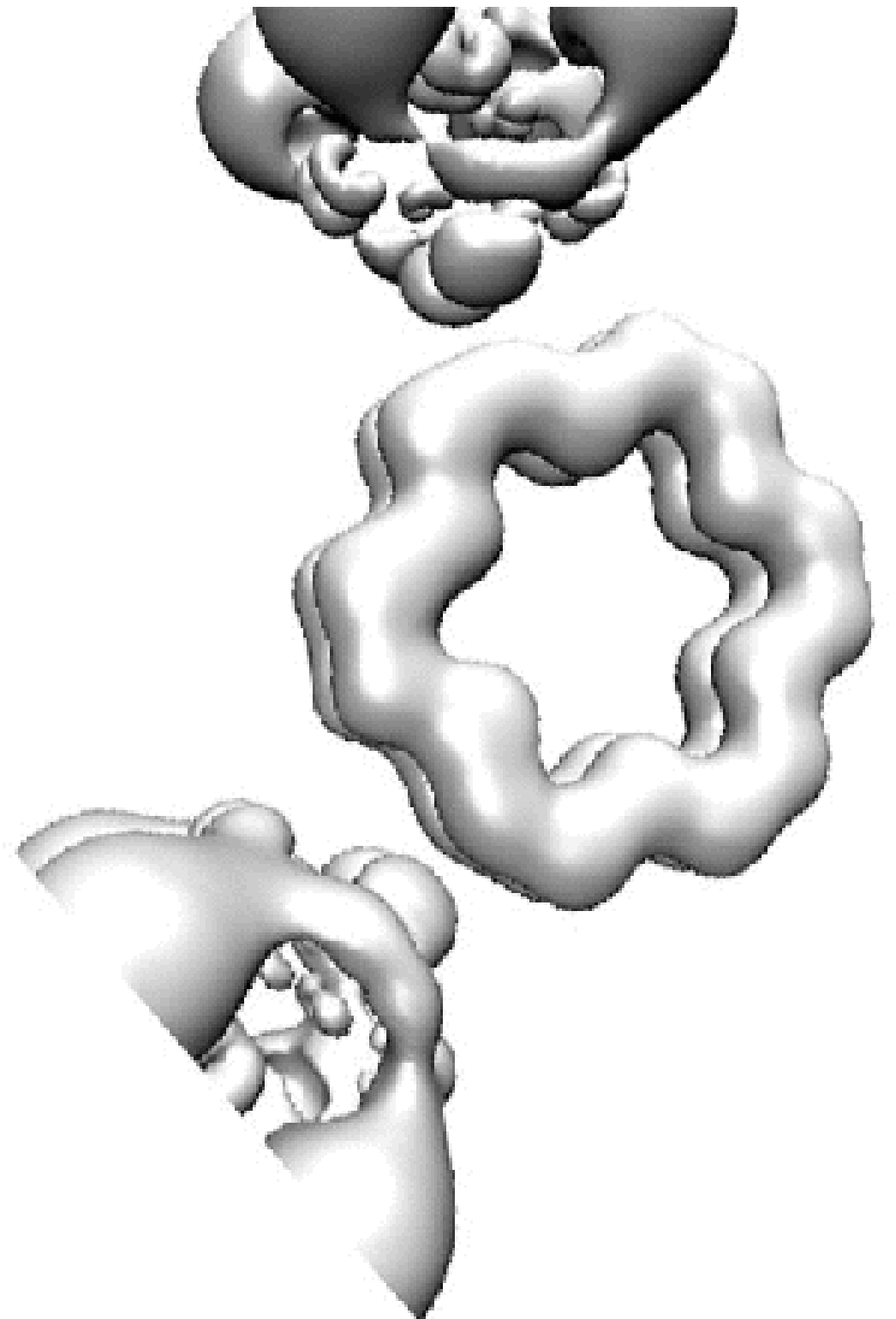}
\caption{
LDOS for the (1,7) [18]annulene/Au system from LDA.
(a) Top view and (b) side view for the energy window [-1.0, -0.2] eV
(HOMO resonance) in Fig.\ 3 (e).
(c) Top view for the energy window [-0.2, 0.2] eV (Fermi energy). 
The $\pi$-orbital signature is clear.
}
\label{fig_LDOS_ann18}
\end{figure}

How to understand the difference in behavior between the small benzene ring
and the large [18]annulene ring? The answer lies in the energies of the
molecular levels. In LDA, for example, the energy of benzene $\pi$
states (HOMO) is -6.5 eV which is in the energy window of the Au-S
bonding at the contact, from -7.0 to -5.7 eV. The larger size of
[18]annulene means that its $\pi$ states (HOMO) are higher in energy, at
-5.0 eV, which is well above the energy window of the Au-S bonding.
Consequently, in [18]annulene, the $\pi$ orbital becomes the frontier
orbital which therefore dominates transport. 

    From this picture one can see that the advantage of using larger
conjugated molecules is two-fold: First, the resulting $\pi$-dominated
transport preserves the simple quantum interference effects. Second, the
longer length suppresses $\sigma$ tunneling, leading to a larger on/off
ratio.


\begin{figure}[tb]
(a)\includegraphics[angle=0,width=3.5cm]{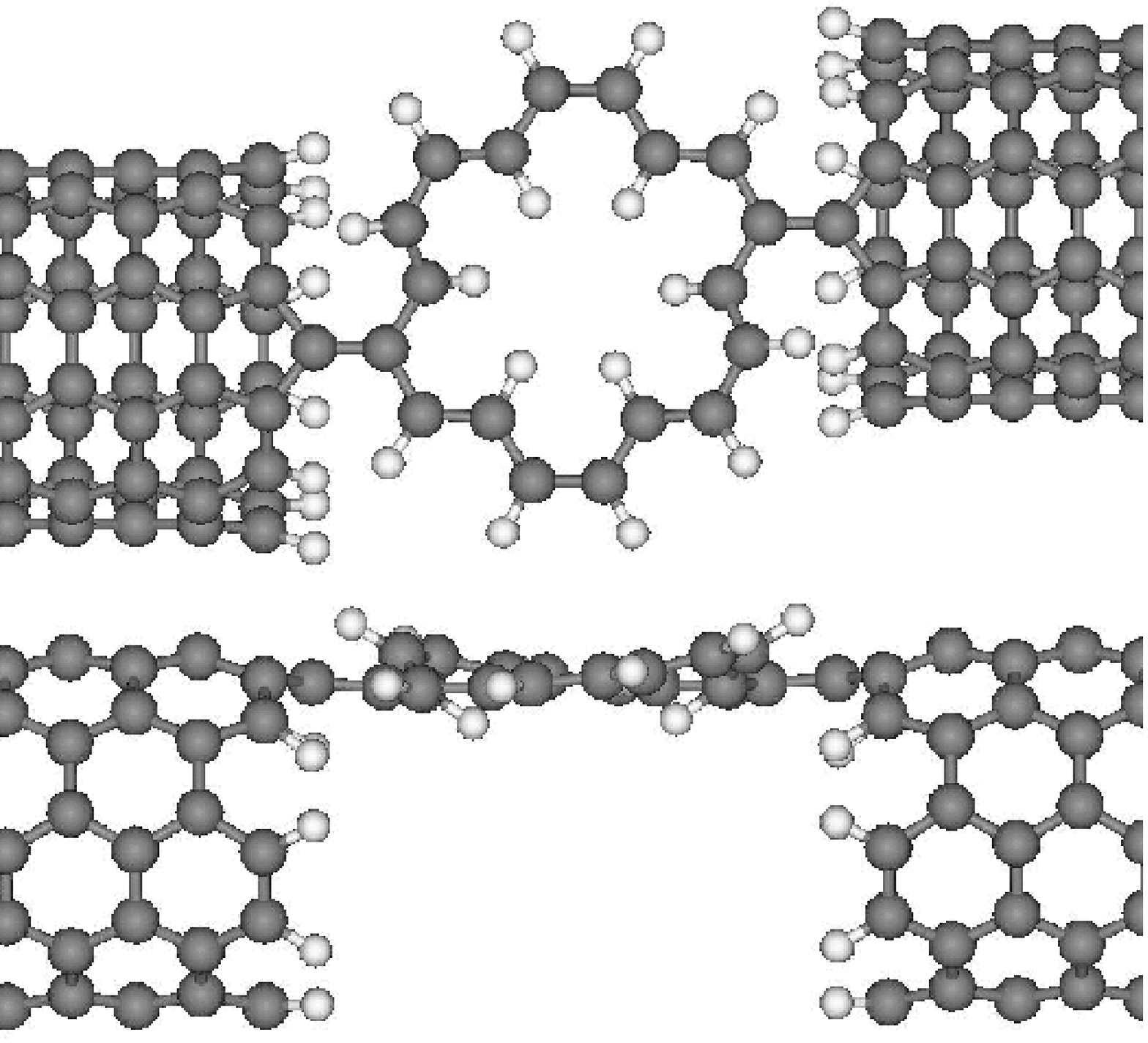}
\includegraphics[angle=0,width=3.0cm]{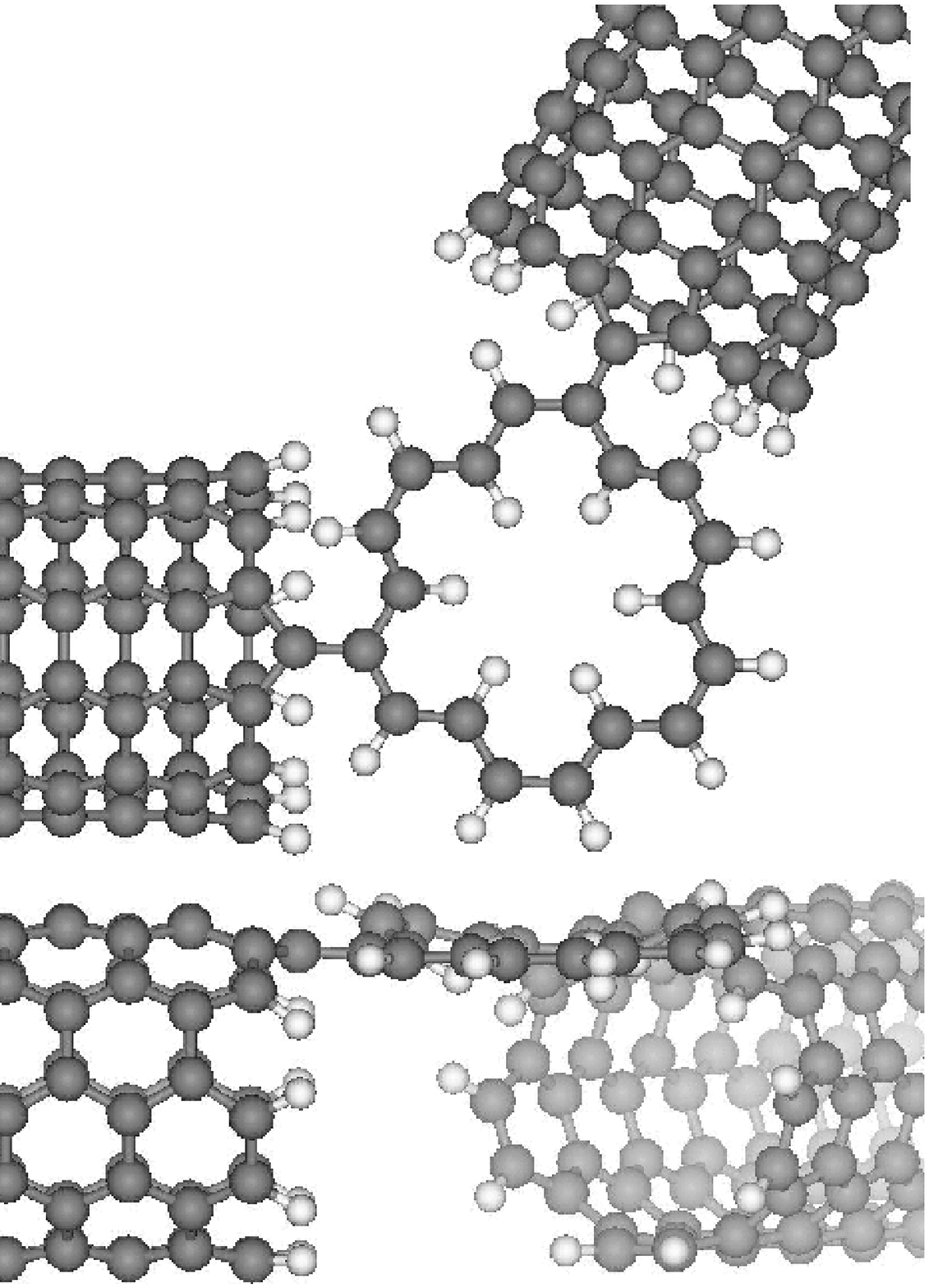}(b)
(c)\includegraphics[angle=0,width=7.5cm]{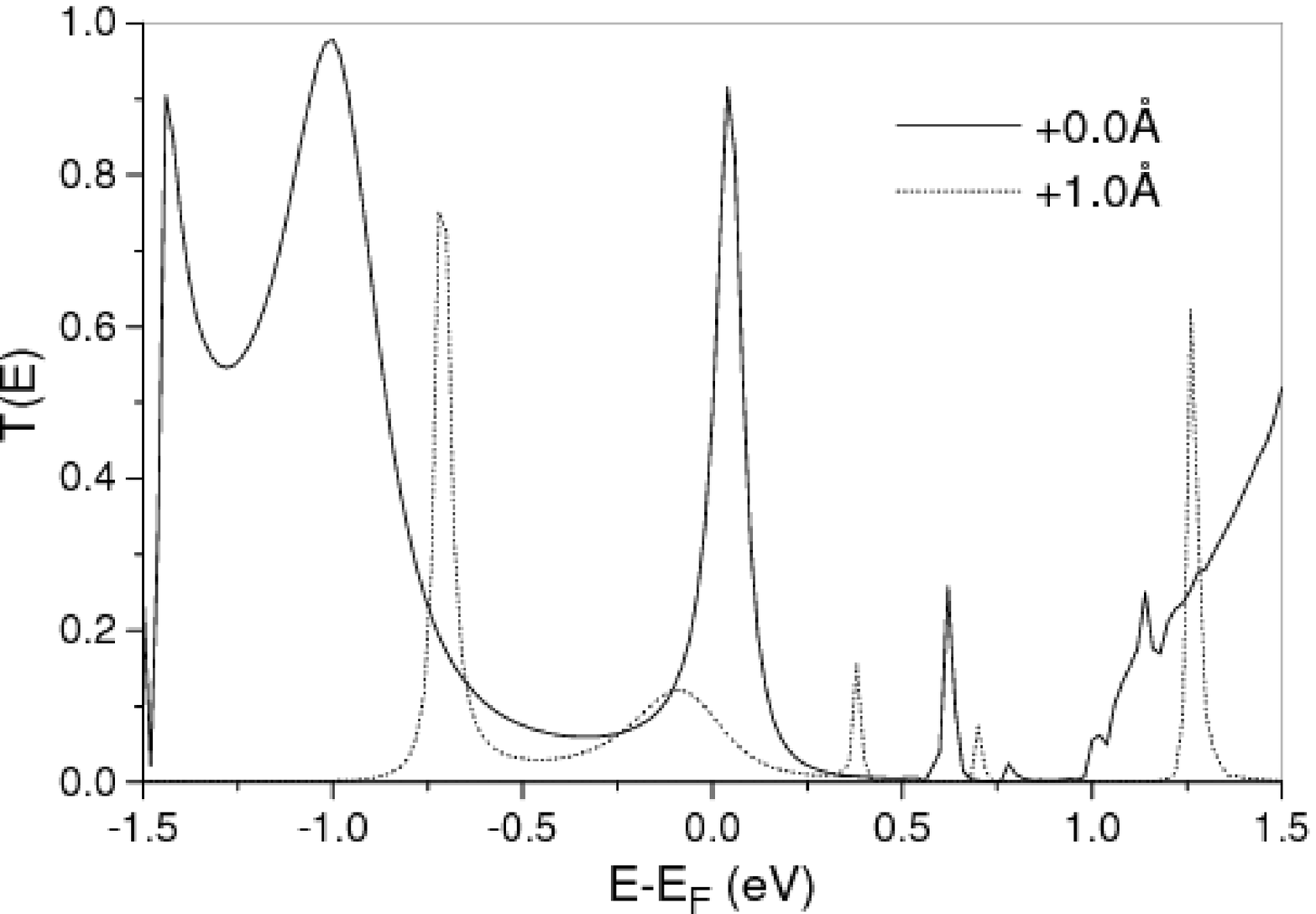}
(d)\includegraphics[angle=0,width=7.5cm]{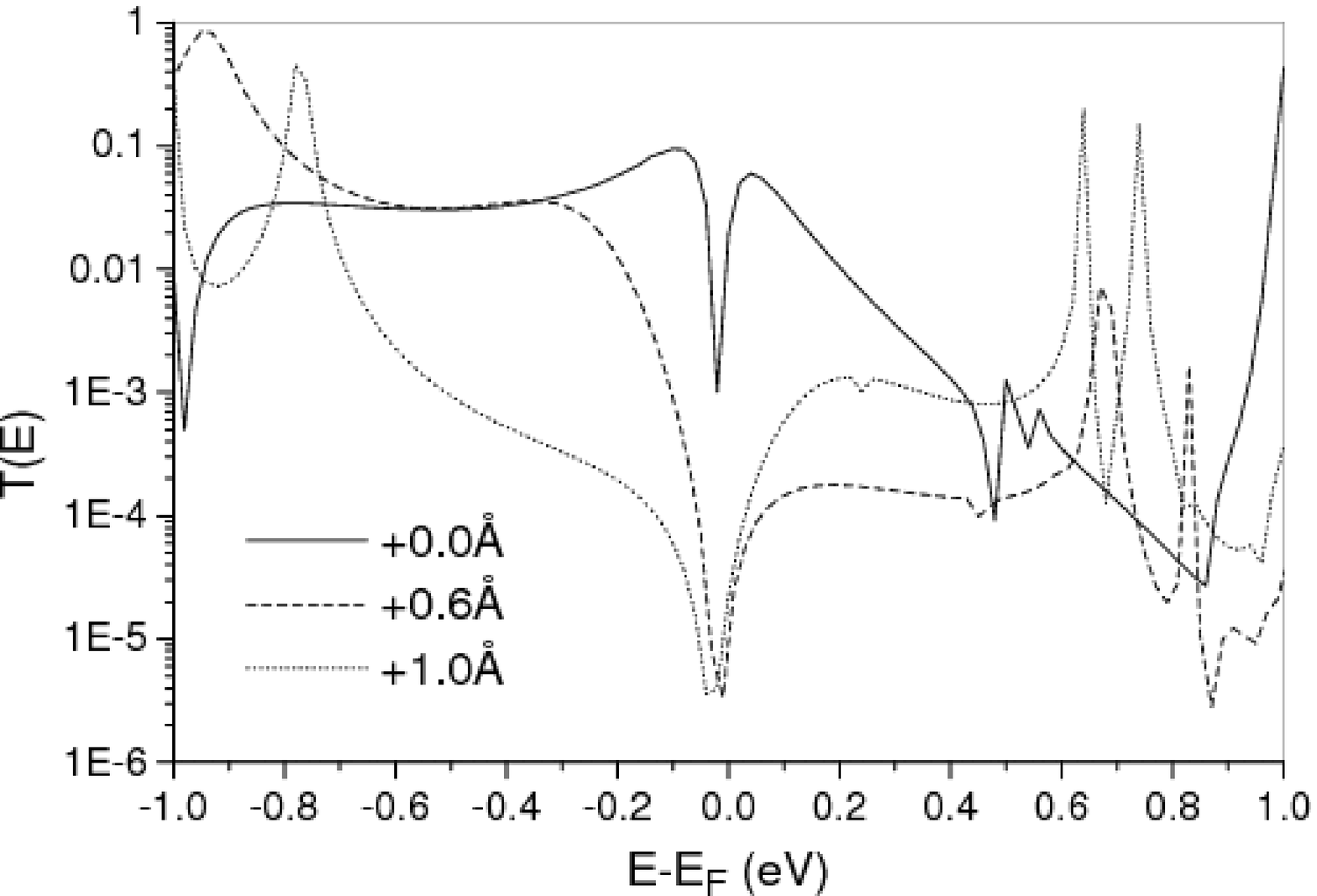}
\caption{
Optimized structures of the 5-member-ring-connected
(5,5)CNT--[18]annulene--(5,5)CNT junctions with the lowest total
energies: (a) constructive (1,10) and (b) destructive (1,7)
configurations. Their transmission functions for different molecule-lead
separations (indicated in the legend) are given in (c) and (d),
respectively. Note the resonance peak around the Fermi energy in (c) and
the anti-resonance peak in (d). 
}
\label{fig_str_55tube}
\end{figure}

While Au electrodes have some advantages, they have a severe
disadvantage as well: the contact transparency is poor (despite the
strong chemical bond), leading to a small equilibrium conductance even
in the case of constructive interference. For a more favorable case with
strong coupling near the Fermi energy, we use a metallic (5,5) carbon
nanotube as a lead, connecting to the molecule through a 5-member-ring
\cite{Ke07CNT-mol}. The optimized structures are shown in Fig.\ 5 for
the (1,10) and (1,7) configurations; both have a co-planar structure
which provides overall conjugation and thus strong molecule-lead
$\pi$-$\pi$ coupling.

\begin{figure}[tb]
(a)\includegraphics[angle=0,width=5.0cm]{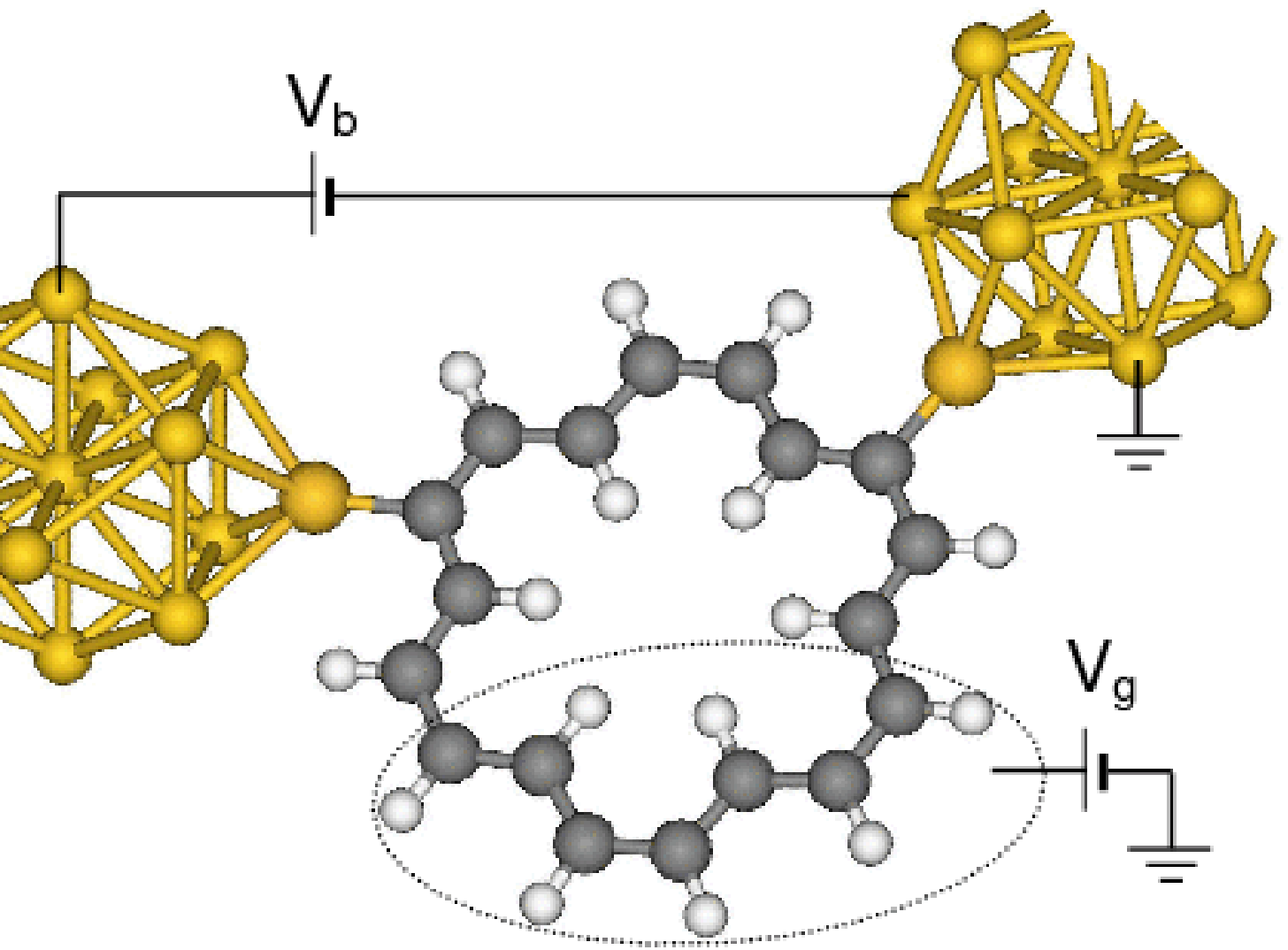}
(b)\includegraphics[angle=0,width=8.0cm]{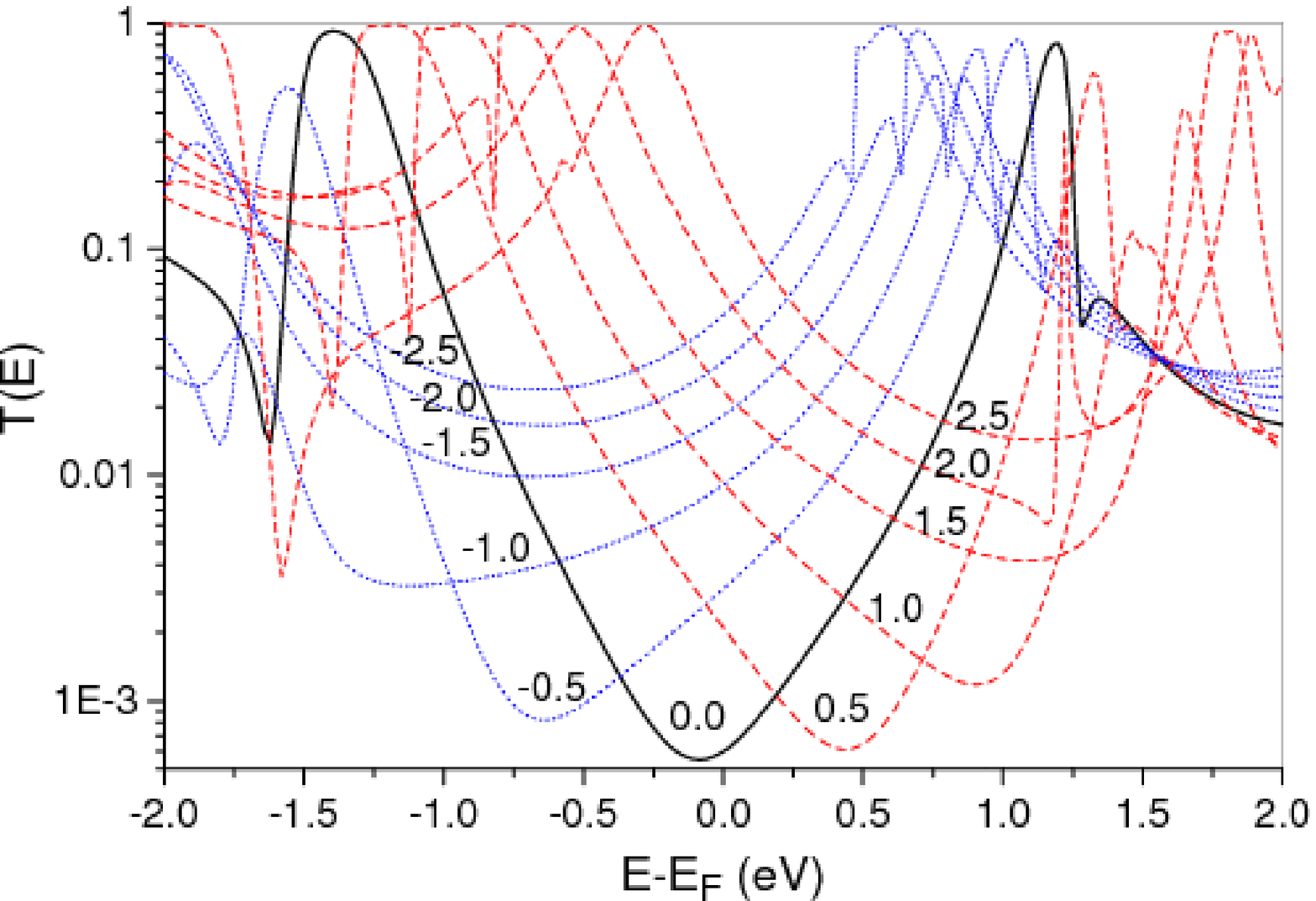}
\caption{
(color online) (a) A schematic drawing of a [18]annulene-based
field-effect QIE transistor. The source-drain current is controlled by
the gate voltage $V_g$ which causes elastic scattering in the gated
region. The transmission functions for zero bias ($V_b$=0) and different
$V_g$ are given in (b).
}
\label{fig_fet_ann18}
\end{figure}

The strong molecule-lead coupling leads to a metal-induced gap state
traveling into the molecule \cite{Ke07CNT-mol}. For short conjugated
molecules, the two states originating from the two leads will meet in
the molecule, causing a resonance peak in $T(E)$ around $E_F$. This is
just what we see in Fig.\ 5(c) for the constructive (1,10)
configuration. [For $T(E)$ here, we do not use the WBL, but rather find
the lead self-energy from an atomistic {\it ab initio} calculation.] The
resonance peak decays as we increase artificially the molecule-lead
separation (weaken the coupling), as shown in Fig.\ 5(c). For the
destructive (1,7) configuration [Fig.\ 5(d)], an \emph{anti-resonance}
peak appears at $E_F$, indicating that the destructive interference
applies in the strong coupling limit to the metal-induced gap state
(though the anti-resonance is rather narrow). For larger molecule-lead
separation (weaker coupling), the anti-resonance becomes wider and
deeper; thus, the molecule-lead coupling affects significantly the
quantum interference but does not destroy it. 
Note that here the metal-induced gap state leads to a much larger
constructive conductance and a larger on/off ratio compared with 
the case of Au leads. 

The effective destructive quantum interference can be used to control
electric current through single molecular devices. To demonstrate this explicitly,
we propose a model device structure, shown in Fig.\ 6, using a
Au-S-[18]annulene-S-Au (1,7) junction, where a local gate potential (due to an STM tip, for instance) is applied to part of the molecular
ring, which will destroy partially the destructive interference because
of the additional scattering, depending on the gate voltage, $V_g$. The
calculation is carried out using the B3LYP functional.
The computational techniques for
transport in the presence of a local gate potential shift has been described
previously \cite{Ke05113401}. Fig.\ 6(b) shows the transmission
functions for different gate voltages. One can see that a small gate
voltage can modify significantly the transmission function of the
device, both shifting and greatly distorting the transmission valley.
Most importantly, a small gate voltage can increase the equilibrium
conductance by up to three orders of magnitude -- a good behavior for
field-effect QIE transistors.

We would like to finish this Letter with a comment about the effect of molecular thermal vibration on QIE. For non-interference-related electron transport through single molecules which are not very long, molecular vibration should have a very small effect on the $I$-$V$ characteristics because the transport is basically ballistic as long as extemely soft vibrational modes (which may lead to comformational changes) are absent. For QIE, the relevant vibrational modes are those which change the length difference between the two paths. These modes therefore involve changing the C-C bond length and  have very high frequencies. As a result, the effect of molecular vibration on QIE for room temperature can be expected to be small.

We thank Charles Stafford for a series of valuable conversations. This work was supported in part by the NSF (DMR-0506953).


\begin{thebibliography}{10}
\bibitem{DZP-optimization}
The structure optimization calculations are performed by using a numerical
  basis set of double zeta plus polarization (DZP) \cite{Soler022745} combined
  with the use of optimized Troullier-Martins pseudopotentials
  \cite{Troullier911993}. The generalized gradient approximation in the version
  of PBE \cite{Perdew963865} is adopted for the exchange and correlation.
\end{thebibliography}

\providecommand{\refin}[1]{\\ \textbf{Referenced in:} #1}

\end{document}